\documentclass[aps,prl]{revtex4}
\usepackage[dvips]{graphicx}
\usepackage{epsfig}
\usepackage{amsmath}
\usepackage{amsfonts}
\usepackage{amssymb}
\usepackage{wasysym}

\def\be{\begin{equation}}
\def\ee{\end{equation}}
\def\bfi{\begin{figure}}
\def\efi{\end{figure}}
\def\bea{\begin{eqnarray}}
\def\eea{\end{eqnarray}}

\begin{document}

\title{Brain modularity controls the critical behavior of spontaneous activity}

\author{R. Russo$^{1}$, H. J. Herrmann$^{2,3}$ and  L. de Arcangelis$^{4}$ }

\affiliation{
$^{1}$ Physics Department, University of Naples 
Federico II, Napoli, Italy\\
$^{2}$  Institute Computational Physics for Engineering Materials, 
ETH, Z\"urich, CH\\
$^{3}$ Departamento de F\'{\i}sica, Universidade Federal do Cear\'a,
60451-970 Fortaleza, Cear\'a, Brazil\\
$^{4}$ Department of Industrial and Information Engineering,
Second University of Naples and INFN Gr. Coll. Salerno, 
Aversa (CE), Italy
}

\begin{abstract}
The human brain exhibits a complex structure made of scale-free highly connected
modules loosely interconnected by weaker links to form a small-world network. 
These  features appear in healthy patients whereas neurological
diseases often modify this structure. An important open question
concerns  the role of brain modularity
in sustaining the critical behaviour of spontaneous activity. 
Here we analyse the neuronal activity of a model, successful in reproducing 
on non-modular networks the scaling behaviour observed in
experimental data, on a modular network implementing the main statistical features
measured in human brain.
We show that on a modular network, regardless the strength of the synaptic
connections or the modular size and number, activity is never fully scale-free.
Neuronal avalanches can invade different modules which results in
an activity depression, hindering further avalanche propagation. Critical
behaviour is solely recovered if inter-module connections are 
added, modifying the modular into a more random structure.
\end{abstract}


\maketitle

One of the crucial questions in biology is the relation between structure and
function: To what extent the particular structure of a living system controls
the performance in specific functions, or is
 pathological behaviour related to deviations from an optimal
structure? This question is of remarkable interest
for what concerns the human brain.
Modularity is the main feature of the brain, composed of functional areas, whose
existence
is well known since more than a century. The brain functional
connectivity network has received wide attention in the literature in recent
years.
In particular the functional network has been found to exhibit scale free properties
\cite{egui}, i.e.,
the absence of a typical connectivity degree, and small world features
\cite{spo}. Experimental results indicate that 
healthy brains exhibit small world features,
 whereas patients affected by neurological diseases are characterized 
by different networks.
In particular, schizophrenic
patients \cite{rubi} or brain tumor patients \cite{bart} exhibit a more random
architecture of the underlying network. 
Within this context, recently Gallos et al \cite{Gal} have developed a detailed analysis of fMRI data
evidencing the complexity of the modular structure of human brain: 
The functional network is composed
of a set of hierarchically organized modules
made of strong links. These modules are
 self-similar structures, not small-world.
However, modules are connected by weaker ties, which make the network
small world, preserving the well-defined modules. 
Remarkably, weak ties are strategically organized in order to maximize information 
transfer with minimal wiring cost \cite{klei,li1,li2}.

Neuronal activity is the fundamental process leading to the complex brain
functions. Even in absence of external stimuli, the brain undergoes a spontaneous
activity which represents about 30\% of the overall activity.
Spontaneous neuronal activity consists of bursts of firing neurons that can last 
from a few to several hundreds of milliseconds and, if analysed at a finer temporal 
scale, exhibit a complex
structure in terms of neuronal avalanches. Indeed,
{\it in vitro} experiments have recorded avalanche activity \cite{beg1,beg2}
from mature cultures of rat cortex, whose 
size and duration distributions follow a power law  with exponents,
$\sim -1.5$ and $\sim -2$, respectively.
This is a typical feature of a system acting in a
critical state, where large fluctuations are present and the response does
not have a characteristic size. The same critical behavior 
has been measured  {\it in vivo} from rat cortical layers 
 \cite{pnas}, from the cortex of
awake adult rhesus monkeys \cite{pnas2}, as well as for
dissociated neurons from rat hippocampus \cite{maz,pas} or leech ganglia \cite{maz}. 
All these experiments record electro-physiological activity by means of microelectrode arrays, 
sampling  small areas of the cerebral system. 
Criticality in brain activity has been also investigated at a larger scale. 
In particular, from fMRI recordings of spontaneous activity in healthy subjects
\cite{frai} the activity correlation length scales with the functional
area size.
Spontaneous brain activity  has been also measured in healthy subjects by 
magneto-encephalography (MEG) \cite{shri} and
found to scale with the  same exponent of neuronal avalanches.
 Experimental data do not exhibit cascades involving the entire
brain, as expected in a system acting in a critical state.

The scaling exponents of neuronal avalanches measured in different experiments are 
quite robust which has stimulated a number of theoretical studies
\cite{zap1,zap2} and neuronal models have successfully 
reproduced the exponents found for neuronal avalanches \cite{br1,lev12,nieb,br4}. 
Several complex networks have been implemented, as well as hierarchical modular
networks \cite{wang}. However, an important open issue is to verify the critical behavior of 
spontaneous activity on networks reproducing closely the human brain at a 
larger scale, as done recently for the role of cortical columns in neural computation \cite{sto}. 
This study addresses the question how the complex human brain structure, and 
in particular modularity (Fig.1),  affects spontaneous activity. 
The problem is of great relevance since spontaneous activity   
plays an important role in evoked activity, i.e., the response of the system to external 
stimulations. 

\section{Results}

We implement a neuronal network model\cite{br1,br2,br3,br5},
able to reproduce the scaling properties of neuronal avalanches on non-modular 
complex networks,
on a structure that has the main statistical properties of the functional network
measured in human brains \cite{Gal} (see Methods).
The aim is 
to enlighten the dependence of the critical features of neuronal
activity on the number $M$, size $N_m$ and average connectivity strength within modules, $S$,
and between modules, $W$. More precisely, we want 
to investigate if the response of the system is really
scale-free, namely if avalanches of all sizes are observed, and to enlighten
the role of modularity features.

Previous results \cite{br1,br2,br4} have shown that the model generates a 
critical avalanche activity, 
 i.e., an avalanche size distribution behaving as a power law over a size range up to the total number 
of neurons. This result is recovered 
on many non-modular complex network structures 
(regular, small world, scale free, fully connected). 
In order to investigate the role of modularity and, more precisely, of the module size 
on the avalanche activity, we first analyse a system where all modules have the same 
number of neurons $N_m$. 
On a modular network the avalanche size distribution exhibits a power law behavior 
only 
for avalanche sizes up to the size of a single module followed by an 
exponential decay 
(Fig.$2$) and increasing $N_m$ leads to the occurrence of larger avalanches. 
The value of the power law exponent is found to be $1.7 \pm 0.11$ by a Kolmogorov-Smirnov 
test with a confidence level 90\%. This value is compatible but slightly larger than the
experimental value for spontaneous activity \cite{beg1} and the numerical value
found for this model on a non-modular network \cite{br1,br5}. This could be due to
the limited scaling regime which is of the order of the single module size. 

Avalanches are not confined within the module where they start. 
Indeed, by direct inspection, we have verified that avalanches can 
reach several modules but
are able to activate only a few neurons in the invaded modules. Avalanches are 
therefore able to carry information to different modules but the invasion of more 
modules results in a 
depression of the overall activity: Because of the weaker inter-module connections, 
activation of a single neuron in a new module leads to a limited activity in  
an environment not previously stimulated by this avalanche. 
Indeed, by rescaling the avalanche size by the number of neurons in 
a module $N_m$ (lower inset Fig.$2$), we notice that the smaller the module size, 
the higher the 
probability to involve in the activity neurons in different modules.
The dependence of the scaling regime on the number of modules is then analysed by 
increasing $M$, keeping $N_m$ fixed (upper inset 
Fig.$2$). 
Interestingly, all distributions for different  numbers of modules
$M$ collapse onto a unique curve: The activity depression effect 
following the activation of more modules is solely controlled by the modular size $N_m$.

Experimental results \cite {Gal} indicate that intra-module connections are  
stronger than 
inter-module ones. We have verified if the average connection strengths plays a 
role in limiting 
the scaling behavior of the avalanche distribution by changing $S$ and
$W$, up to $W=S$ (Fig.$3$). 
As expected, at fixed $N_m$ and $W$, larger avalanches become more probable for 
increasing $S$ (black line and
triangles).  Conversely, by progressively increasing
$W$ at fixed $S$ we observe that surprisingly the scaling regime decreases and 
the distributions extend to smaller avalanche sizes 
(curves with the same symbol). Indeed 
stronger inter-module connections make the invasion of different modules 
more probable, with a consequent
depression of the overall activity as discussed above. Therefore, confining activity
within a single module makes larger avalanches more probable. 
However, this effect becomes less relevant for increasing modular size 
$N_m$ and avalanches involving the entire system are never observed even for $W=S$, 
independently of their value (red triangles and green line), 
suggesting that the only relevant parameter is the ratio $W/S$. 
For large $N_m$ the exponent value is independent of $S$ and $W$
and $W$  solely affects the onset of the exponential cut-off.

In real brains external stimulation often involves more functional regions. 
We have monitored the system 
activity under the simultaneous stimulation of several neurons in different modules. 
By increasing the number of stimulated neurons chosen at random in different modules, 
the size of the largest avalanche increases, however activity is never able to 
involve 
all neurons even in the limiting case $W=S$ (Fig.$4$). Conversely, under 
stimulation of several 
neurons a lack of small avalanches is observed, which further limits the scaling 
regime of the avalanche 
size distribution. As a final verification, we consider a network made of modules 
with different sizes (Fig.$4$). The presence of small and large modules 
increases the occurrence probability of large avalanches, 
i.e., events involving about half system are observed. However, the scaling regime is 
again controlled by the 
size of the largest module. Interestingly, in all cases the value of the power law 
exponent is independent of the features of the modular network.

In order to better investigate the role of modularity on the avalanche activity, 
we monitor 
the number of modules $m$ involved in each avalanche and measure the distribution $P(m)$ 
(Fig.5). This exhibits a power law behavior $P(m)\sim m^{-\mu}$ over a scaling range 
increasing
with the total number of modules in the network $M$, followed by an exponential cut-off.
In the upper inset we show that the scaling $P(m)=m^{-\mu}f(m/m_{\rm max})$ is satisfied,
with the maximum number, $m_{\rm max}$, of modules reached by activity scaling with the number
of modules in the network, $m_{\rm max}\propto M^{\sigma}$ with $\sigma=0.70
\pm 0.05$.
As expected, for fixed $S$ the probability to invade more modules increases for larger $W$ 
and for $W\lesssim S$ avalanches reaching all modules can be observed. As a consequence 
the scaling behavior depends on $W$ for fixed $S$. More precisely, the power law exponent 
is solely dependent on the ratio $\beta =W/S$ (lower inset Fig.5) and exhibits a  value 
$\mu=3.6 \pm 0.1$ for $\beta>0.5$. 
The value of the exponents $\mu$ and $\sigma$ are obtained by a Kolmogorov-Smirnov
test with a confidence level 95\%.
The distribution crosses over towards an exponential 
behavior for smaller $\beta$.
As observed in real brains,
inter-module connections being slightly, but not significantly, weaker than connections within 
modules improve the functional efficiency of the system, since activity can
 reach all modules optimizing information transmission.

Results indicate that neither the module size, nor the number of modules, the stimulation 
extension or the connection strengths can make the modular network recover a truly critical 
behavior for avalanche activity. Modularity in the network is insured by the very different 
percentage of intra-module (92\%) and inter-module (8\%) connections. Increasing this last
percentage would progressively modify and finally destroy the 
modular feature of the network. Indeed, by increasing the number of inter-module connections 
the avalanche size distribution finally recovers a power law behavior over the entire 
scaling range (Fig.6) if connections are equally distributed within and between
modules. This effect is observed even more clearly for longer plastic
adaptation, namely for a wider distribution of synaptic strengths (red straight line).
Conversely, for modular networks (8\% of inter-module connections) the
distribution is not affected by the synaptic strengths 
and the two distributions for different connection strengths collapse
(black symbols and grey straight line). Moreover, for increasing inter-module
connections avalanches have a higher probability to invade more modules (inset of Fig.6) 
and for 50\% inter-module connections avalanches invade with  equal
probability all modules.

\section{Discussion}

Modular organization is a characteristic feature of many biological systems, especially 
relevant in systems exhibiting separate functional units.
In neuronal networks, modularity plays a crucial role in network synchronization and 
affects the synchronization transition
more than long-range connections \cite{boc}. Hierarchical modular structure has been also 
found able to
sustain scale-free activity in presence of weak perturbations \cite{wang}. 
However, how the modular structure
affects the scale free behaviour is an open question which can provide interesting insights in
the origin of pathological response in some neuronal systems.
The present study shows evidence that the modular structure of the brain, made of highly 
connected functional areas loosely inter-connected by weaker links, is a fundamental 
ingredient insuring efficient and controlled functioning of healthy brains. Modularity, 
at the same time, allows information to reach all areas but hinders the involvement of the 
whole system in activity, which could lead to pathological response. 
This behaviour is due to an activity depression resulting from the invasion
through weaker links of
modules not stimulated by previous firings, which tend to confine the
response to the modular size. These results are in agreement with experimental
data, since the critical behavior 
of brain activity is confirmed at the level of  single functional areas,
i.e., non-modular networks with different connectivity properties. At the scale of the entire 
brain, conversely, our results explain why critical behaviour should not be observed, 
regardless the modular size, the number 
of modules and the connectivity strengths. The only way 
to recover criticality is to increase inter-modular connectivity,
transforming the modular network and
leading to the loss of modularity of 
the functional network, which becomes a more random structure, 
as in patients with neurological diseases. This
suggests that the structure could be at the
origin of such pathologies. Results are in 
agreement with a study of the stability of the power grid in US via the sand-pile 
model  on a two-module network \cite{bru}. 
Also in this case modularity mitigates large 
cascades by diverting load at an optimal percentage of inter-module connections close to 8\%.

{\bf METHODS}

{\bf Implementation of the modular network.}
Here we implement a single level of the modular network measured in ref. \cite{Gal} (Fig.1).
We consider $N$ neurons uniformly distributed in a 2$d$ system.
Neurons are organized in $M$ modules each containing $N_m$ neurons.
The simulated values of $M$ are comparable with the number of functional areas or
Brodmann's area in the brain, under the requirement that $N_m$ is large enough 
to provide a satisfactory scaling regime for the size distributions.
Both, the case of modules with the same and different size are analysed. Synaptic
connections and strengths $g_{ij}$ are assigned according to ref. \cite{Gal},
with $g_{ij}\neq g_{ji}$.
More precisely, connections between neurons within the same module form a scale-free network,
where the out-going connectivity degree
is distributed as $n(k_{out})\propto k_{out}^{-\gamma}$ with $\gamma=2.1$, and
their
initial value $g_{ij}$  is assigned at random in the interval $[S-0.1, S+0.1]$.
The two neurons are chosen according to a distance dependent probability,
$p(r)\propto e^{-r/5<r>}$, where $r$ is their spatial distance and $<r>$ a characteristic
distance \cite{roe}.  Results do not depend on the neuronal
spatial density or the average connection length  $<r>$. 
Since the synaptic connections between modules represent 8\% of the total number of
connections \cite{Gal}, we evaluate their number on the basis of the total number of
intra-module connections as $Nk_{inter}=Nk_{intra}0.08/0.92$.
We then assign them according  to
$P(k_{out}^i, k_{out}^j)\sim (k_{out}^i)^{1-\gamma}(k_{out}^j)^{-\epsilon}$, where
$k_{out}^i$ and $k_{out}^j$ are the number of synapses of neurons $i$ and $j$
belonging to different modules and  $\epsilon=2.1$, according to the experimental value. This implies
that inter-module connections are typically established between low connectivity neurons and
hubs are connected with neurons within the same module.
Each inter-module connection has an
initial random strength $g_{ij}$ in $[W-0.1, W+0.1]$, where $W<S$
according to ref. \cite{Gal}. The total out-connectivity degree distribution is updated during
each step of the procedure, to include both inter-module and intra-module connections.
Once the network of output connections is established, we identify the
resulting degree of in-connections, $k_{in}$ for each neuron.

\smallskip
{\bf Neuronal network model.}
We consider $N$ neurons characterized by their potential
$v_i$, initially assigned at random. The initial distribution is irrelevant in order
to obtain critical avalanche activity. Each neuron can be
excitatory or inhibitory, with a percentage of $p_{in}=10\%$
inhibitory synapses. This value, lower than the percentage measured
experimentally, is needed in order to observe a sufficient scaling regime in avalanche
size distributions. As it will be shown in the Results section, the scaling regime is
controlled by the single module rather than the entire system size, therefore a larger
 $p_{in}$ would require to simulate networks with larger $N_m$.
Moreover, inhibitory neurons are chosen among neurons with high connectivity ($k_{out}>50$),
according to experimental observations \cite{bon}.
Each neuron fires as soon as its potential reaches a threshold $ v_{\rm max}$.
In order to start activity a small random stimulus (about  $v_{\rm max}/10$)
is applied to a random neuron. As soon as at a given time the
value of the potential at a neuron $i$ is above threshold,
$v_i \geq v_{\rm max}$, the neuron fires modifying the potential
of the $k_{out_i}$ connected neurons
proportionally to $g_{ij}$ \cite{br5},
$v_j(t+1)=v_j(t)\pm \frac{ v_ik_{out}^i}{k_{in}^j}\frac{g_{ij}(t)}{\sum_k g_{ik}(t)}$,
where the sum over $k$ is over all out-going
connections of $i$ and the plus or minus sign
denotes excitatory or inhibitory $g_{ij}$, respectively.
The ratio $k_{out}^i/k_{in}^j$
 makes the potential variation of neuron $j$ induced by neuron $i$ independent of the
connectivity level of both neurons, as expected for real neurons \cite{br5}.
After firing, a neuron is set to a zero resting potential and
 in a refractory state lasting one time step
(about 10 ms), during which it is unable to  receive or transmit any charge.
At the end of an avalanche, we implement Hebbian plasticity rules:
The strength of synapses connecting active neurons is increased proportionally to
the activity of the synapse, i.e., the
potential variation of the post-synaptic neuron,
$g_{ij}(t+1) =g_{ij}(t) +(v_j(t+1)-v_j(t))/v_{\rm max}$.
Conversely, the strength of all inactive synapses is
reduced by the average strength increase per synapse,
$\Delta g = \sum_{ij, t} \delta g_{ij} (t)/ N_b$,
where $N_b$ is the number of synapses.
The same strengthening and weakening rules are applied to both
excitatory and inhibitory synapses.
The presence of both strengthening and weakening rules implements a
homeostatic regulatory mechanism for synaptic strengths, which
preserves the average synaptic strength \cite{roy}.
An external stimulus then triggers further activity in the system.
We implement  plasticity during a finite series of stimuli in order to
modify the synaptic strengths initially assigned at random. However, we verify that 
plastic adaptation preserves the average values of the strength distribution, 
in particular that inter-module connections
are weaker than intra-module ones. 
Moreover, we limit the duration of plastic adaptation to avoid pruning 
of synaptic strengths, which would modify the network modular structure.

\begin{figure}
\includegraphics[width=6.5cm]{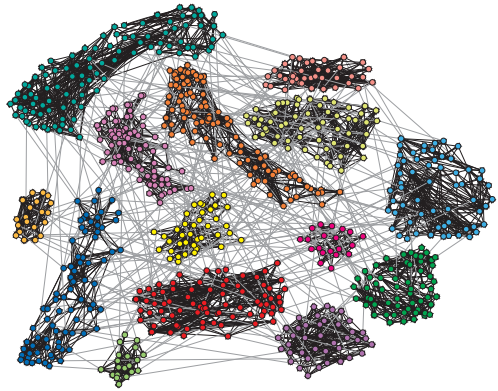}
\caption{{\bf Modular structure of the neuronal system.}
Network with $M=14$ modules in different
colors with a number of neurons from $N_m=20$ to 100.
Intra-module connections have a larger
average strength (thicker lines) than inter-module connections, i.e. $S>W$.
For the network generation procedure see Methods.
}
\label{fig1}
\end{figure}

\begin{figure}
\includegraphics*[width=6.5cm]{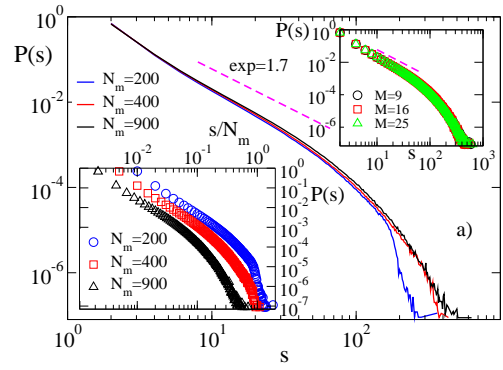}
    \caption{{\bf Dependence of the scaling behaviour of the avalanche size
distribution on system size.} Avalanche size distribution for 100 configurations 
of modular networks with $M=25$, $W=0.2$, $S=0.4$ and
for different values of $N_m$.
The dashed line has a slope $1.7$.
Lower inset: The same curves plotted as function of $s/N_m$.
Upper inset: Distributions evaluated for $N_m=900$ and different module numbers
$M$.
     }
    \label{fig2}
\end{figure}

\begin{figure}
\includegraphics*[width=6.6cm]{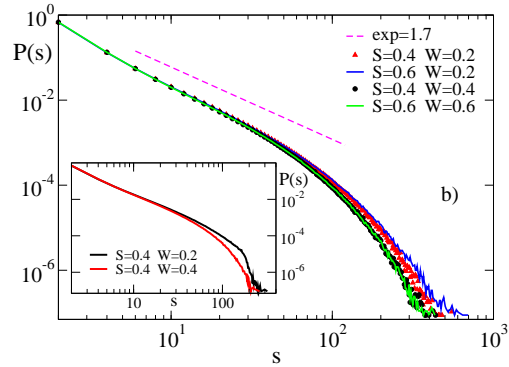}
    \caption{{\bf Dependence of the scaling behaviour of the avalanche size
distribution on synaptic strengths.} Avalanche size distribution
for 100 configurations of modular networks with $M=16$, $N_m=900$ and different values 
of $W$, $S$. The dashed line has a slope $1.7$.
Inset: Distributions for  a smaller module size $N_m=200$.
     }
    \label{fig3}
\end{figure}

\begin{figure}
\includegraphics*[width=7cm]{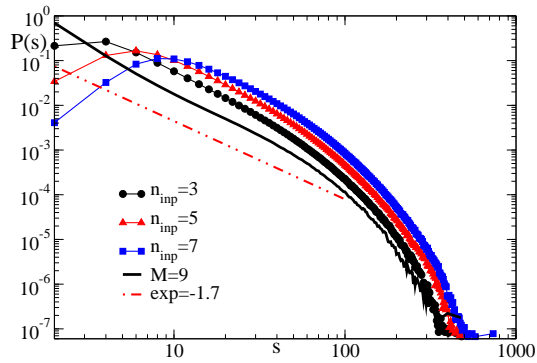}
    \caption{{\bf Scaling behaviour for multiple stimulation and different module size.}
Avalanche size distribution
for 100 configurations of modular networks where $n_{inp}$ neurons are
initially stimulated. Parameter values are
$M=16$, $N_M=400$, $W=S=0.4$ (symbols).
Avalanche size distribution for networks with different $N_m$
($M=9$, $S=0.4$, $W=0.2$, $N_m=100, 400, 900$).
The dot-dashed line has a slope $1.7$.
     }
    \label{fig4}
\end{figure}

\begin{figure}
\includegraphics*[width=7cm]{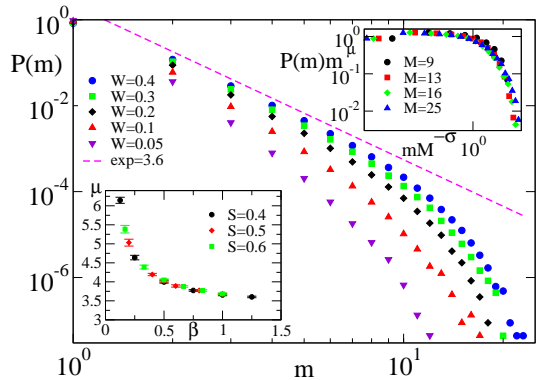}
    \caption{{\bf Scaling properties of the distribution of modules invaded by avalanches.}
Distribution of the number of modules reached by an
avalanche for 100 configurations of modular networks with $M=25$, $N_m=900$
for different values of $W$ and fixed $S=0.4$.
The dashed line indicates the power law decay with an exponent $\mu=3.6$.
Lower inset: The power law exponent $\mu$ as function of $\beta=W/S$ for different $W$ and $S$.
Upper inset: Universal scaling function obtained by rescaling of the distributions
$P(m)m^{\mu}$ versus $mM^{-\sigma}$.
     }
    \label{fig:fig5}
\end{figure}

\begin{figure}
\includegraphics*[width=7cm]{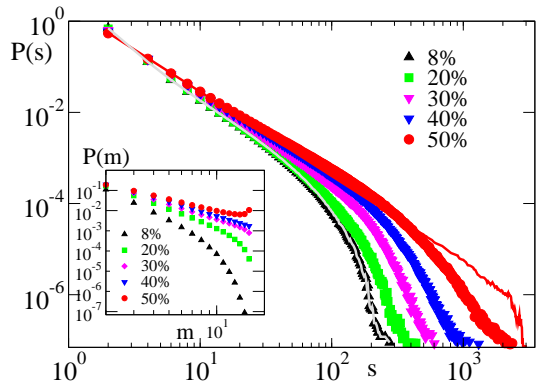}
    \caption{{\bf Destruction of modularity leads to recovery of full scale free behaviour.}
Avalanche size distribution for 100 configurations of
networks with $M=16$ and $N_m=200$
neurons and for increasing percentage of inter-module connections ($W=S=0.6$)
and limited plasticity (symbols). Straight lines show the same
distributions for 8\% and 50\%  and longer plastic adaptation (grey and
red lines respectively).
Inset: Distributions $P(m)$ for the same networks.
     }
    \label{fig:fig6}
\end{figure}

{\bf Acknowledgments.}
We acknowledge financial support from the
ERC Advanced Grant 319968-FlowCCS and the SNF project 205321-13874.

{\bf Author contributions}
L.d.A. and H.J.H. were involved in all the phases of this study. 
R.R. performed the numerical simulations and prepared the figures.
All authors reviewed the manuscript.

{\bf Competing financial interests:} 
The authors declare no competing financial interests.

\end{document}